\documentclass[aps,prl,twocolumn]{revtex4}
\usepackage{amssymb}

\usepackage{graphicx}

\begin{document}

\title{Fermi-surface-topology tuned high temperature
superconductivity in the cuprates}
\author{N.~Harrison, R.D.~McDonald, and J.~Singleton}
\affiliation{National High Magnetic 
Field Laboratory, Los Alamos National Laboratory, MS E536,
Los Alamos, New Mexico 87545}
\date{\today}

\begin{abstract}
Based on recent magnetic-quantum-oscillation,
ARPES, neutron-scattering and resistivity data,
we propose that the extraordinarily high 
superconducting transition temperatures
in the cuprates are driven by an exact mapping of the
incommensurate spin fluctuations onto the $d_{x^2-y^2}$
Cooper-pair wavefunction. 
Strong evidence for this comes from the
direct correspondence between the incommensurability factor $\delta$
seen in inelastic neutron scattering and the inverse superconducting
coherence length (see Fig.~2).
Based on these findings, one can specify the
optimal characteristics of a solid 
that will exhibit ``high $T_{\rm c}$'' superconductivity.
\end{abstract}

\maketitle
\begin{figure}[htbp!]
\centering
\includegraphics[width=0.45\textwidth]{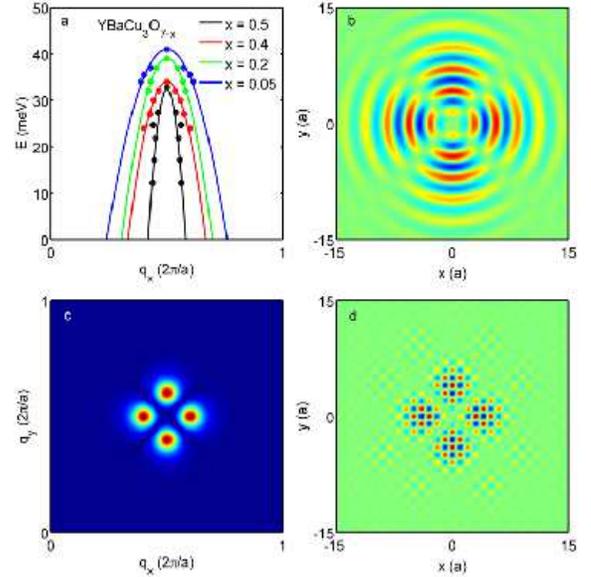}
\caption{(a)~Energy ($E$) versus $q_x$ for several 
YBa$_2$Cu$_3$O$_{7-x}$ compositions~\cite{dai1,stock1}, 
with fits to Eq.~\ref{dispersion} to
obtain extrapolated values of $\delta$ at 
$E=$~2~meV (dotted line). (b)~Contour plot of $d$-wave Cooper-pair wavefunction
in two-dimensions (Eq.~\ref{cooper}); we assume a Fermi 
wavevector $k_{\rm F}=\pi/\sqrt{2}a$ where 
$a$ is the lattice parameter, while $p=0.1$ for 
which we select $\xi_0=$~27~\AA~from Table~\ref{table1}. 
(c)~Simulation of the incommensurate neutron scattering peaks 
obtained by Fourier transforming Eq.~\ref{spinfluct};
$\delta$ and $\xi$ correspond to $p=0.1$ (Table~\ref{table1}).
(d)~Real-space spin-fluctuation map
corresponding to the set of four incommensurate peaks
in (b), also for $p=0.1$ (see Table~\ref{table1}). We consider the simple case where $s_{\bf Q}$ is the same for all ${\bf Q}$.}
\label{fig1}
\end{figure}

In spite of strong evidence for an order parameter
exhibiting $d_{x^2-y^2}$ symmetry~\cite{tsuei1} in the
``High $T_{\rm c}$'' cuprate superconductors,
experiments establishing a direct link with the
pairing mechanism have not been forthcoming.
However, with the advent of magnetic-quantum-oscillation
measurements on underdoped cuprates~\cite{leyraud1,yelland1,sebastian1} 
and consequent proposals for the
Fermi-surface topology that are consistent with 
experiments ranging from neutron scattering to
ARPES~\cite{julian1,harrison1,sebastian1},
there is the opportunity to resolve this issue.
We find that the $d_{x^2-y^2}$ Cooper-pair
wavefunction is exactly tuned to the real-space
form of the fluctuations responsible for
the incommensurate antiferromagnetic scattering
observed in inelastic neutron experiments
on all doped 
cuprates~\cite{kampf1,schrieffer1,yamada1,dai1,stock1,wakimoto1} 
(see e.g. Fig.~\ref{fig1}a). The
dispersion of these
fluctuations, which are likely to be driven by the
varying topology of the Fermi surface as
the hole density $p$ increases~\cite{sebastian1}, 
provides a very plausible 
energy scale for 
$T_{\rm c}$ in the cuprates~\cite{schrieffer1}.

Most treatments of superconductivity
consider the state as a 
condensate~\cite{cooper1,bardeen1,tinkham1,lee1,kadin1};
in such a picture, it is natural to treat
the Cooper-pair wavefunction in $k$-space.
However, we are interested in the similarity of the
spatial distribution of spin-fluctuations
and the Cooper pair, and so
we consider the form of the
two-dimensional 
$d_{x^2-y^2}$ Cooper-pair wavefunction in
real space~\cite{tsuei1,kadin1}:
\begin{equation}
\psi({\bf r}) \propto \cos(rk_{\rm F}) (x^2-y^2)
{\rm e}^{-3r/\xi_0}.
\label{cooper}
\end{equation}
Here $r= \sqrt{x^2+y^2}$ is the cylindrical polar
radius, $x$ and $y$ are corresponding
Cartesian coordinates, $k_{\rm F}$ is the 
Fermi wavevector~\cite{leyraud1,yelland1}
and $\xi_0$ is the superconducting coherence length.
Figure~\ref{fig1}(b) shows a contour plot of
$\psi({\bf r})$; the known diagonal
nodal regions~\cite{tsuei1,lee1} are clearly visible.
In such a plot, the lengthscale over
which $\psi({\bf r})$ is non-negligible
is defined by $\xi_0$ 
(Eq.~\ref{cooper}).
We now consider similarities
between this wavefunction and the
real-space topology of the incommensurate
antiferromagnetic fluctuations seen in
inelastic neutron-scattering~\cite{kampf1,schrieffer1,yamada1,dai1,stock1,wakimoto1}.

Low-energy inelastic neutron-scattering data
for both La$_{2-x}$Sr$_x$CuO$_4$~\cite{yamada1,wakimoto1} and
YBa$_2$Cu$_3$O$_{7-x}$~\cite{dai1,stock1} follow an 
approximately inverted parabolic form
\begin{equation}
E({\bf q}) = E_0\left(
1-\frac{a^2({\bf q}-{\bf Q}_0)^2}{\pi^2\delta^2}
\right),
\label{dispersion}
\end{equation}
where $a$ is the in-plane lattice parameter and 
$E_0$ is a doping-dependent 
energy scale~\cite{kampf1,schrieffer1,yamada1,dai1,stock1};
some examples are shown in Fig.~\ref{fig1}(a).
Here, $\delta$
depends on the composition and doping of the
cuprate involved~\cite{kampf1, schrieffer1,monthoux1};
some values
are given in Table~\ref{table1}.
As $E\rightarrow 0$, the brightest scattering intensity 
typically appears at a cluster of four incommensurate
peaks 
at ${\bf Q} = (\pm \pi/a, \pm(1\pm2\delta)\pi/a)$
and $(\pm(1\pm2\delta)\pi/a, \pm \pi/a)$~\cite{kampf1,schrieffer1,yamada1,dai1,stock1,wakimoto1};
a simulation is shown in the
positive $q_x$, positive $q_y$ quadrant
of two-dimensional $k$-space in Fig.~\ref{fig1}(c).
The form of the scattering peaks in Fig.~\ref{fig1}(c)
is already very
suggestive of the $d_{x^2-y^2}$ wavefunction
in Fig.~\ref{fig1}(b); to make a quantitative
comparison, however, we require a real-space representation
of the corresponding spin fluctuations.

\begin{figure}[htbp!]
\centering
\includegraphics[width=0.5\textwidth]{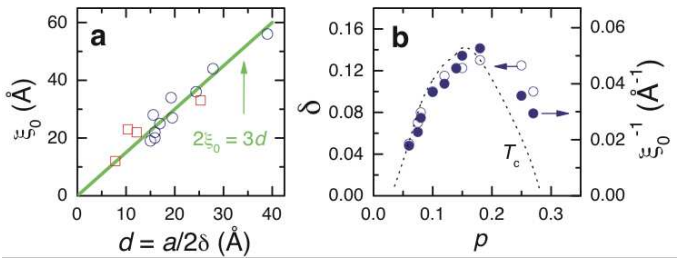}
\caption{{\bf a} The approximately linear correspondence between 
$d=a/2\delta$ and $\xi_0$ for La$_{2-x}$Sr$_x$CuO$_4$ ($\circ$) 
and YBa$_2$Cu$_3$O$_{7-x}$ ($\square$). 
The line represents the ratio $\xi_0/d=$~3/2. 
{\bf b} Plot showing $\delta$ and $\xi^{-1}_0$ in 
La$_{2-x}$Sr$_x$CuO$_4$ exhibiting a maximum versus 
$p$ in a similar fashion to $T_{\rm c}$ 
(scaled for comparison), though $T_{\rm c}$ falls 
more rapidly with $p$ beyond optimal doping.} 
\label{fig2}
\end{figure}

A simple model of spin fluctuations
that can produce the observed incommensurate
scattering peaks
is a sinusoidal variation of the staggered
moment modulated by an exponential damping factor;
the latter term represents the fact that 
the antiferromagnetic
fluctuations possess a finite correlation length $\xi$~\cite{kampf1,monthoux1}.
The spatially-varying moment is thus
\begin{equation}
s({\bf r},t) = \sum_{\bf Q}s_{\bf Q}\exp(-\frac{r}{\xi}+{\rm i}\omega t)
\cos({\bf Q}.({\bf r}-{\bf r}_0)),
\label{spinfluct}
\end{equation}
where $\omega$ is the angular frequency of the fluctuations,
${\bf r}_0 = (\pm d,0)$ or $(0,\pm d)$
with $d=a/2\delta$, and
the sum in {\bf Q} runs over the values given above.
The simulation shown in Fig.~\ref{fig1}(c) is
a Fourier transform of Eq.~\ref{spinfluct}
that reproduces the general form of the experimental 
neutron data~\cite{kampf1,schrieffer1,yamada1,dai1,stock1,wakimoto1}
very well. Note that the
choice of ${\bf r}_0$ is quite critical;
any other value would give significant
intensity at ${\bf q} = {\bf Q}_0$, at
variance with the experimental spectra.
This is an important point, to which we will return.
Finally, Fig.~\ref{fig1}(d) shows a 
time-averaged contour plot
of Eq.~\ref{spinfluct}. The similarity between the
spin fluctuation distribution and
the $d_{x^2-y^2}$ wavefunction (Fig.~\ref{fig1}(b))
is most marked: both show similar angular 
and radial distributions. Less obvious,
but equally germane, is the phase
of each function plotted. The choice of
${\bf r_0}$ necessary to reproduce the
neutron data means that there is a $\pi$ difference
of phase between adjacent lobes of the spin-fluctuation
distribution, just as there is a $\pi$ difference
in phase between adjacent lobes of the
$d_{x^2-y^2}$ Cooper-pair wavefunction~\cite{tsuei1,lee1}.

\begin{figure}[htbp!]
\centering
\includegraphics[width=0.45\textwidth]{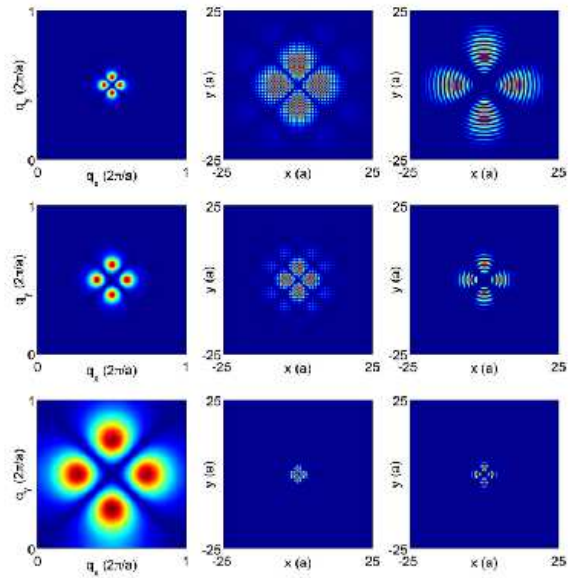}
\caption{Comparisons of the spin density amplitude and 
$d_{x^2-y^2}$ wavefunction at selected values of $p$. 
The first column shows plots of the incommensurate 
diffraction peaks obtained on re-Fourier transforming 
$s({\bf r})$ calculated using Eq.~\ref{spinfluct} and the published 
values of $\delta$ and $\xi$ listed in Table~\ref{table1}. 
The second column shows the calculated $|s({\bf r})|$ 
while the third column shows $\rho_{\rm c}({\bf r})=|\Psi({\bf r})|^2$ 
calculated according to the values of $\xi_0$ listed in 
Table~\ref{table1}. 
Rows 1, 2 and 3 consecutively correspond to 
$p\approx$~0.06, 0.1 and 0.152.} 
\label{fig3}
\end{figure}

As mentioned above,
the lengthscale of the Cooper-pair wavefunction
is $\xi_0$; the corresponding lengthscale for
the spin fluctuation distribution (Eq.~\ref{spinfluct})
is $d=a/2\delta$~\cite{kampf1,schrieffer1,yamada1,dai1,stock1}.
To further establish
the causal relationship between the spin fluctuations
and the superconducting pairing mechanism, we now compare
independent estimates of these lengthscales
for several cuprates.
The values of $d$ are obtained from inelastic
neutron-scattering experiments. In 
the La$_{2-x}$Ca$_x$CuO$_4$ cuprates,
the dispersion relationships (Fig.~\ref{fig1}(b))
have been measured down to
low energies, providing accurate values of $\delta$~\cite{yamada1}.
However, there is a loss of intensity at lower
energy transfers in YBa$_2$Cu$_3$O$_{7-x}$,
possibly because of their greater homogeneity~\cite{dai1,stock1},
necessitating a downward extrapolation of the
dispersion curves to obtain $\delta$ in the limit
$E\rightarrow 0$. The coherence lengths $\xi_0$ are independently
derived from magnetoresistance and other data~\cite{ando1,wen1};
occasionally some interpolation is required
to obtain values for the same hole doping
and composition as the neutron-scattering
data. Fig.~\ref{fig2}a shows the
corresponding values of $\xi_0$ plotted
against $d$. Table~\ref{table1} lists the values
(plotted against doping for La$_{2-x}$Sr$_x$CuO$_4$ in Fig.~\ref{fig2}b).

Remarkably, Fig.~\ref{fig2}a shows
that the $\xi_0$ versus $d$
values for {\it all} compositions and dopings
listed in Table~\ref{table1} lie on a single
straight line $\xi_0/d=$~3/2, irrespective of whether
La$_{2-x}$Sr$_x$CuO$_4$ or YBa$_2$Cu$_3$O$_{7-x}$
is considered, suggesting a common origin for
the spin fluctuations and the
superconductivity in all of the cuprates.

Having established the experimental relationship between
$\xi_0$ and $d$, we can now plot the
evolution with increasing hole density
$p$ of the spatial distribution
of the spin fluctuations (Eq.~\ref{spinfluct})
alongside the corresponding Cooper-pair
wavefunction (Eq.~\ref{cooper}) using experimental
values of both parameters; Fig.~\ref{fig3} shows the result.
At small values of $p$, the incommensurate
neutron-scattering peaks occupy a small
area of $k$-space~\cite{yamada1,dai1,stock1};
the corresponding real-space
spin fluctuation distribution occupies a large area,
as does the Cooper-pair wavefunction.
As $p$ increases, the incommensurate
peaks spread out in $k$-space~\cite{yamada1,dai1,stock1}; hence,
the spin-fluctuation spatial distribution
is compressed, as is the Cooper-pair wavefunction.

\begin{table}[h!b!p]
\caption{Parameters for various cuprate 
superconductors, including hole doping $p$. 
For La$_{2-x}$Sr$_x$CuO$_4$, $\delta$ values measured 
at neutron transfer energies of $E\approx$~2-3~meV 
are taken from Refs.~\cite{yamada1,wakimoto1}. 
For YBa$_2$Cu$_3$O$_{7-x}$, $\delta$ values 
at the equivalent energy are taken from 
plots such as Fig.~\ref{fig2}a using data from 
Refs.~\cite{dai1,stock1}. 
Correlation lengths $\xi$ are taken 
from Ref.~\cite{kampf1} while BCS coherence 
lengths $\xi_0$ are taken from Refs.\cite{wen1,ando1}, 
occasionally requiring interpolation between 
adjacent compositions for precise 
doping matches to $\delta$.}
\begin{tabular}{llllll}\\
\hline
compound & $p$ & $\delta$ & $\xi$~(\AA) & $\xi_0$~(\AA) & $T_{\rm c}$~(K) \\
\hline
La$_{1.94}$Sr$_{0.06}$CuO$_4$ & 0.06 & 0.05 & 17 & 56 & 13\\
La$_{1.925}$Sr$_{0.075}$CuO$_4$ & 0.075 & 0.07 & 14 & 44 & 33\\
La$_{1.92}$Sr$_{0.08}$CuO$_4$ & 0.08 & 0.08 & 11 & 36 & 24\\
YBa$_2$Cu$_3$O$_{6.5}$ & 0.082 & 0.078 & 9 & 33 & 59 \\
YBa$_2$Cu$_3$O$_{6.6}$ & 0.097 & 0.16 & 7.5 & 22 & 62.7 \\
La$_{1.9}$Sr$_{0.1}$CuO$_4$ & 0.10 & 0.10 & 12 & 27 & 29\\
YBa$_2$Cu$_3$O$_{6.8}$ & 0.123 & 0.18 & 4 & 24 & 71.5 \\
La$_{1.88}$Sr$_{0.12}$CuO$_4$ & 0.12 & 0.115 & 17 & 25 & 33.5\\
La$_{1.86}$Sr$_{0.14}$CuO$_4$ & 0.14 & 0.1225 & 12 & 22 & 35\\
La$_{1.85}$Sr$_{0.15}$CuO$_4$ & 0.15 & 0.1225 & 14 & 20 & 38\\
YBa$_2$Cu$_3$O$_{6.95}$ & 0.152 & 0.24 & 3.3 & 12 & 93 \\
La$_{1.82}$Sr$_{0.18}$CuO$_4$ & 0.18 & 0.13 & 9 & 19 & 35.5\\
La$_{1.75}$Sr$_{0.25}$CuO$_4$ & 0.25 & 0.125 & 8 & 28 & 15\\
La$_{1.73}$Sr$_{0.27}$CuO$_4$ & 0.27 & 0.1 & 10 & 34 & 7\\
\hline
\end{tabular}
\label{table1}
\end{table}

Before turning to the origin of
the incommensurate scattering peaks in
the neturon data~\cite{kampf1,schrieffer1,yamada1,dai1,stock1,wakimoto1,sebastian1},
it is worth summarizing
two factors that lead us to propose a causal
relationship (or one might say spatial resonance)
between superconductivity and antiferromagnetic
fluctuations in the cuprates.\\
{\it First}: on setting ${\bf r}_0=(\pm d,0)$ or $(0,\pm d)$ 
where $d=a/2\delta$,
the $\pi$ difference in phase between the 
adjacent `spin clusters' in Fig.~\ref{fig1}d is 
aligned with the $\pi$ difference in phase between 
adjacent lobes of the $d_{x^2-y^2}$ Cooper-pair wavefunction.
The spin fluctuations are therefore ``in phase'' with the
Cooper-pair wavefunction.\\ 
{\it Second}: independent of our choice of wavefunction, 
there exists a very direct correspondence between 
$\xi_0$, which determines the overall size of the Cooper pair, 
and $\delta$, which quantifies the physical separation 
$d=a/2\delta$ between adjacent clusters of spins in 
real space (see Fig.~\ref{fig2}). 
On combining both of these observations, 
the $\xi_0/d=$~3/2 slope in Fig.~\ref{fig2}a 
can be seen to originate from a direct spatial 
overlap between the vector locations 
$(\pm d,0)$ and $(0,\pm d)$ at which the spin 
fluctuation amplitude is largest and those 
$(\pm\frac{2}{3}\xi_0,0)$ and $(0,\pm\frac{2}{3}\xi_0)$ 
at which the probability amplitude of the $d_{x^2-y^2}$ 
wavefunction given by Eqn. (\ref{cooper}) is largest.
While there are certainly systematic and random errors 
in the methods used to extract $\xi_0$ values~\cite{wen1,ando1}, 
they are consistent in La$_{2-x}$Sr$_x$CuO$_4$ with estimates 
obtained from strong-coupling variants of 
the BCS formulae $\xi_0=\hbar v_{\rm F}/\pi\square_0$ 
and $2\square_0/k_{\rm B}T_{\rm c}\approx 4-5$,
where $\square_0$ is the $T=0$ order 
parameter~\cite{bardeen1,tinkham1}, 
using typical Fermi velocities 
$v_{\rm F}\approx$~6-8~$\times$~10~$^4$~ms$^{-1}$ determined 
from quantum oscillation experiments on YBa$_2$Cu$_4$O$_8$ 
and YBa$_2$Cu$_3$O$_{6.5}$~\cite{yelland1,leyraud1}. 
This provides a natural explanation for the observed
linear dependence of $\delta$ on $T_{\rm c}$~\cite{yamada1,wakimoto1}. 
A slightly stronger coupling $2\square_0\sim/k_{\rm B}T_{\rm c}$~6 
is required for the YBa$_2$Cu$_3$O$_{7-x}$ series~\cite{dai1,stock1}.

Incommensurate diffraction peaks in metallic systems
are almost always associated with Fermi-surface nesting,
in which the periodicity of spin or change modulation
is determined by the topology of the Fermi surface~\cite{gruner1}.
Whilst the tight-binding calculations of the cuprate
Fermi surface usually give a single, 
large hole pocket~\cite{julian1,harrison1},
the recent experimental observations of (multiple) small pockets
in the underdoped cuprates~\cite{leyraud1,yelland1,sebastian1}
supports the suggestion 
that some form of nesting occurs~\cite{julian1,harrison1,sebastian1}.
Added weight is given by the 
similarity of the incommensurate mode dispersion 
(Fig.~\ref{fig1}(b)) to those observed in itinerant 
antiferromagnets such as Cr and 
V$_{2-x}$O$_3$~\cite{bao1,stockert1,endoh1} 
(both above and below the Ne\'{e}l temperature). 
In such a picture, the cuprate Fermi surface plays an 
increasingly important role as the hole doping $p$ is increased. 
Thus, just as in V$_{2-x}$O$_3$~\cite{bao1}, 
large-moment antiferromagnetic insulator behavior in the 
cuprates eventually gives way to small moment 
incommensurate itinerant antiferromagnetic behavior 
as the system becomes more
metallic~\cite{harrison1,kampf1}. 
Consistent with itinerant magnetism, the orbitally-averaged 
Fermi velocity 
$v_{\rm F}=\sqrt{2e\hbar F}/m^\ast\approx$~8~$\times$~10~$^4$~ms$^{-1}$,
where $F$ is the Shubnikov-de Haas oscillation frequency,
of the pockets in YBa$_2$Cu$_3$O$_{6.5}$~\cite{leyraud1} 
is comparable to the mode velocity 
$v_0=2aE_{\rm r}/\pi\hbar\delta\approx$~14~$\times$~10$^4$~ms$^{-1}$ 
(at $E=$~0) that one obtains fitting Eqn.~\ref{dispersion} 
to the $E$-versus-${\bf q}$ data points obtained from 
inelastic neutron scattering experiments on the same 
sample composition~\cite{stock1,note2} (Fig.~\ref{fig1}(b)).
As $p$ increases toward optimum doping, $E_0$ in Fig.~\ref{fig1}(a) also
increases~\cite{kampf1,schrieffer1,monthoux1}, providing
a suitable characteristic energy scale ($\sim 10$s~meV, 
{\it i.e.} $\sim 100$s of Kelvin) for 
$T_{\rm c}$~\cite{schrieffer1}.

In conventional BCS superconductors, the quasiparticle
interactions that result
in pairing are via charge coupling
to acoustic phonon modes~\cite{bardeen1,tinkham1}.
The incommensurate spin fluctuations have a
dispersion relationship (Fig.~\ref{fig1}(a))
analogous to the acoustic phonons ({\it i.e.}
approximately linear as $E\rightarrow 0$, saturating at
a maximum energy that defines 
the energy scale of $T_{\rm c}$).
The remaining ingredient in the problem is
therefore a coupling mechanism
between the spin fluctuations and the charge
inherent in the Cooper pair.
This is found in the large onsite correlation energy $U$,
which inhibits double occupancy of spins 
or holes~\cite{zhang1}. Consequently, local variations 
in the density of holes 
$\Delta\rho_{\rm h}$ and spin-density amplitude 
$\Delta s$ are expected to be subject to the relation 
\begin{equation}
\Delta\rho_{\rm h}\propto-\Delta|s|.
\label{reciproc}
\end{equation} 
Pairing mechanisms involving this behavior have been 
considered both in the weak (small Hubbard $U$)~\cite{schrieffer2} 
and strong (large Hubbard $U$)~\cite{mott1} coupling 
limits, although typically in conjunction with long-range 
antiferromagnetic order. However, there is no reason to suspect
that such mechanisms will not apply in
regimes where the antiferromagnetism is strongly 
fluctuating~\cite{kampf1}. 
Therefore, because of the reciprocity relationship
(Eq.~\ref{reciproc}), the slowly varying modulation 
of the spin fluctuation intensity should be accompanied 
by a concomitant charge modulation
$\bar{\rho}_{\rm h}({\bf r})\propto-|\bar{s}({\bf r})|$. 
The very similar form of
${\rho}_{\rm h}({\bf r})\propto-|{s}({\bf r})|$
in the second column of Fig.~\ref{fig3} to 
$\rho_{\rm c}\propto|\Psi({\bf  r})|^2$ 
in the third column of Fig.~\ref{fig3} 
therefore provides direct evidence for a `spatial charge 
commensurability' between the Cooper pair 
wavefunction and incommensurate spin fluctuations.

Qualitatively, the Cooper pairs in this work
are spatially compact near optimum doping (Fig.~\ref{fig3}),
resembling the strong coupling spin-bipolarons 
of Mott and Alexandrov~\cite{mott1,zhang1}. 
Away from optimum doping,they become spatially extended, like a 
weak-coupling $d_{x^2-y^2}$ variant of the 
`spin bags' proposed by Schrieffer 
{\it et al.}~\cite{schrieffer2}. 
The difference between these pictures is that the fluctuations
mediating the superconductivity in the present proposal
are characteristic of a system on the brink of long-range
order, rather than an established antiferromagnet. 

In summary, based on measurements
including Fermi-surface studies~\cite{leyraud1,yelland1}, 
neutron scattering data~\cite{kampf1,schrieffer1,yamada1,dai1,stock1}
and resistivity experiments~\cite{ando1,wen1}, we propose that the unusually high superconducting transitions
in the cuprates are driven by an exact mapping of the
incommensurate spin fluctuations onto the $d_{x^2-y^2}$
Cooper-pair wavefunction. The spin fluctuations are
driven by the Fermi-surface topology,
which is prone to nesting~\cite{julian1,harrison1};
they couple to the itinerant holes via the strong on-site
Coulomb correlation energy $U$, which prohibits double
occupancy of spins or holes~\cite{zhang1}. The maximum energy of the
fluctuations ($\sim 100$s of 
Kelvin~\cite{kampf1,schrieffer1,monthoux1}) gives  an appropriate
energy scale for the superconducting $T_{\rm c}$.  
Based on these findings, one can specify the
features necessary for
a solid to exhibit ``high $T_{\rm c}$'' superconductivity;
(i)~the material should be quasi-two-dimensional,
with the conducting layers exhibiting four-fold symmetry
(to ensure that fluctuations are optimally configured to
a $d$-wave order parameter);
(ii)~the material should have a Fermi-surface topology
susceptible to nesting, so as to produce antiferromagnetic 
fluctuations with a large degree of incommensurability $\delta$;
(iii)~however, the electron-phonon coupling should be moderate,
to prevent formation of stripe or
charge-density-wave-like phenomena  that compete
with supercoductivity~\cite{tranquada1}; in  other oxides,
the larger electron-phonon coupling dominates,
preventing superconductivity~\cite{peter1}.

We are grateful to Ed Yelland,
Roger Cowley, Andrew Boothroyd, Wei Bao, Sasha Balatsky
and Bill Buyers
for stimulating discussions.
This work is supported by the
US Department of Energy (DoE) BES
program ``Science in 100~T''.
NHMFL is funded by
the National Science
Foundation, DoE and the State of Florida.


\begin{thebibliography}{99}
\bibitem{note2}
Ideally, $v_0$ should be compared with the Fermi velocity of the unreconstructed Fermi surface, which is expected to be somewhat higher owing to its larger cross-section in $k$-space~\cite{julian1,harrison1}.
\end{thebibliography}
\end{document}